\documentclass[aps,reprint,pre,twocolumn]{revtex4-1}%
\usepackage{amsmath}
\usepackage{amsfonts}
\usepackage{amssymb}
\usepackage{graphicx}
\usepackage{bbold}
\usepackage[colorlinks,linkcolor=blue,anchorcolor=blue,citecolor=blue,urlcolor=black]%
{hyperref}
\usepackage{mathrsfs}
\usepackage{dcolumn}
\usepackage{bm}
\usepackage{epsfig}
\usepackage[version=3]{mhchem}%
\def\beq{\begin{equation}}
\newcommand{\eeq}{\end{equation}}
\newcommand{\ee}[1] {\label{#1} \end{equation}}
\newcommand{\pS}{\ensuremath{{\cal M}}}          
\newcommand{\refeq}  [1] {(\ref{#1})}
\newcommand{\bea}{\begin{eqnarray}}

\newcommand{\eea}{\end{eqnarray}}
\newcommand{\Lop}{\ensuremath{{\cal L}}}       
\begin{document}
\title{Thermodynamics of chaotic relaxation processes}
\author{Domenico Lippolis}
\email{domenico@ujs.edu.cn}
\affiliation{School of Mathematical Sciences, Jiangsu University, Zhenjiang 212013, China}
\date{\today}

\begin{abstract}
    
The established thermodynamic formalism of chaotic dynamics, 
valid at statistical equilibrium, is
here generalized to systems out of equilibrium, that have yet to relax to a steady state. 
A relation between information, escape rate, and the phase-space average of
an integrated observable  (e.g. Lyapunov exponent, diffusion coefficient) is obtained for finite time. Most notably, the thermodynamic treatment may predict the phase-space profile of any integrated observable for finite time, from the leading and subleading eigenfunctions of the Perron-Frobenius/Koopman transfer operator. Examples 
of that equivalence are shown, and the theory is tested analytically on the 
 Bernoulli map, while numerically on the perturbed cat map, the H\'enon map, and the Ikeda map,
 all paradigms of chaos. 

\end{abstract}

\maketitle

\section{Introduction}
The exponential stretching and folding of phase-space densities that characterizes 
 chaotic dynamics makes long-time evolution unpredictable, and with that the problem of motion
intractable. It is then customary to study the statistical properties of the 
phase space, and in particular to aim at estimating long-time expectation values of relevant observables,
under the assumption of asymptotic relaxation of the system to an equilibrium- or a stationary state.

In this framework  finds its roots 
the thermodynamic formalism~\cite{Rufus75,BeckSchl,Gaspbook,Ruelle04},  developed from the 1970s on,
that is based on the idea of using large-deviation theory~\textcolor{red}{\cite{Touch09}} to define a dynamical analog of the thermodynamic
Gibbs states, which, at statistical equilibrium, do maximize the generating functional of the desired averages, their fluctuations, and multitime correlation functions. This approach
is at the basis of the formulation of evolution operators and periodic orbit theory~\cite{dasBuch}, and it
has more recently been employed to elucidate the relation between Lyapunov exponents and decay of 
correlations~\cite{Slip13}, as well as to
identify dynamical phase transitions in deterministic chaos\textcolor{red}{~\cite{Naftali,Espig,Monthus}}.  

This paper aims at extending the thermodynamic formalism
 of chaotic dynamics to out-of-equilibrium systems. The Gibbs states of the original formulation
 are here generalized to include time-dependent weights, later identified with phase-space densities
 transported by the transfer operator (Perron-Frobenius or Koopman) that governs the time evolution
 of the system. The so-surmised probabilities for the dynamical microstates give rise to a time-dependent
 free energy (`topological pressure'), that is related not only to the finite-time dynamical averages
 of interest to us, but crucially, to the entire phase-space profiles of the relevant observables. 

 In the remainder of the present section, I will concisely review the fundamentals and the main results of the
thermodynamic formalism of chaotic systems at statistical equilibrium.  In section~\ref{FiniteTime}, I shall
extend the key definition of Gibbs probability for an integrated observable on a chaotic trajectory to
a system that has yet to reach equilibrium.  Consequently, I derive an expression for finite time, which relates escape rate and R\'enyi information with free energy (`topological pressure') and a 
newly introduced quantity named tilted information. This non-equilibrium first law 
approaches the renowned Kantz-Grassberger 
relation~\cite{KantzGrass85} as the system relaxes to equilibrium or a stationary state. Section~\ref{AvSec} contains the
main implications of the out-of-equilibrium thermodynamic formalism: any finite-time integrated observable is expressed in terms of the first two eigenfunctions of the Perron-Frobenius- or Koopman operator in a way that its phase-space profile does not depend on the choice of the observable.
These claims are validated analytically on the Bernoulli shifts,
a paradigm of chaos (Sec.~\ref{BerSec}), and numerically on: $i)$
the perturbed cat map (Sec.~\ref{CAT}), a hyperbolic system
defined on a torus, which has no escape and whose transient dynamics is entirely ruled by the
second eigenfunction of the transfer operators; $ii)$ the Hamiltonian H\'enon map (Sec.~\ref{HamHen})
with weak additive noise,
whose dynamics is essentially governed by a chaotic saddle, but it also features an isolated, marginally stable
fixed point, which makes the system non-hyperbolic locally. 

While the dependence of the out-of-equilibrium observables on the first two eigenfunctions of the
transport operator is general, the universality of their profiles breaks down when squeezing is introduced.
It is the case of strange attractors, where distinct integrated observables exhibit different phase-space 
dependence. This is exemplified in Sec.~\ref{SecIkeda} by the noisy Ikeda attractor.

\subsection{Gibbs states}
In the thermodynamics of equilibrium, the entropy $S$ is maximal when the internal energy $E$ is fixed (microcanonical ensemble),
while the free energy $F$ is minimized when the internal energy fluctuates (canonical ensemble)  
\begin{equation}
F  = E - k_BT\,S
\,. 
\label{Fenergy}
\end{equation}
The Gibbs (canonical) ensemble is made of a number of subsystems, each occurring with a probability $p_j$: 
\begin{eqnarray}
E &=& \sum_j p_j\,E_j \\
S &=& -\sum_j p_j\,\ln p_j
\,,
\end{eqnarray}
so that ($\beta=1/k_BT$)
\beq
F =  \sum_j p_j\,E_j + \frac{1}{\beta}\sum_j p_j\,\ln p_j
\,.
\eeq
Using Calculus, one can show that the free energy is minimized by choosing~\textcolor{red}{\cite{Chand}}
\begin{equation}
p_j = \frac{e^{-\beta\,E_j}}{\sum_i e^{-\beta\,E_i}}
\,.
\end{equation}
Here $p_j$ is the probability of the subsystem labelled by $j$ to have energy $E_j$.
At equilibrium,
\begin{equation}
F_{\mathrm{min}} = -\ln\,\sum_i e^{-\beta\,E_i}
\,.
\end{equation}

The Gibbs formalism has been used to describe chaotic dynamics of ergodic and mixing systems,
which, in general do not reach thermodynamic-, but rather statistical equilibrium, asymptotically.
Here goes a summary of how that works.

\subsection{Thermodynamics of chaos}
The analysis that follows is formulated for low-dimensional chaotic systems, such as expanding maps on the interval, or hyperbolic Axiom A systems,
such as the cat map, whose phase space may be partitioned to a number of distinct regions, and
trajectories can be encoded with symbolic sequences~\cite{dasBuch}, tracking the regions 
visited by each orbit at every instant.
In the thermodynamical picture of chaotic dynamics, the phase space is the whole (canonical) ensemble,
whose subsystems/Gibbs states are identified with the single (ergodic) trajectories, the latter thought of as infinite sequences in the long-time limit.    

In the formalism, the Boltzmann constant $\beta$ becomes a parameter:
given the map $x_{t+1}=f(x_t)$, that defines the dynamics,
identify the energy with the 
integrated observable
\beq
E_j(t) = A^t_j = \sum_{\tau=0}^t a\left[f^\tau(x_0)\right]
\,,
\ee{E_intobs}
where 
$j$ tags the symbolic sequence of the trajectory that starts at $x_0$ and is iterated by $\tau$ times by $f$, and $a$ is in general a function of an operator, for example the differential, in the evaluation of the Lyapunov exponents, or the squared position to yield 
the diffusion constant as the integrated observable~(\ref{E_intobs}).  
Assuming that the system reaches equilibrium (or a stationary state) for $t\rightarrow\infty$,
it has been proven~\cite{Rufus75,BeckSchl} 
that the Gibbs probabilities 
\beq
p_j(t) = \frac{e^{-\beta\,E_j(t)}}{\sum_i e^{-\beta\,E_i(t)}}
\label{GibbsP}
\eeq
minimize the `free energy', or equivalently, maximize the quantity
\beq
{\cal{P}}(\beta) = \lim_{t\rightarrow\infty} \frac{1}{t}\ln \sum_j e^{-\beta\,E_j(t)} 
\,,
\label{TopP}
\eeq
that is known as `topological pressure'~\cite{footnote1}, although it is in fact a free energy.
It has been shown that Eqs.~(\ref{GibbsP})-(\ref{TopP}) imply the relation  
\begin{equation}
{\cal{P}}(\beta) = \langle A\rangle_\beta + h_\beta
\,,
\label{PAhEq}
\end{equation}
analogous to Eq.~\refeq{Fenergy} for the free energy, where $\langle\cdot\rangle$ denotes an
ensemble average, and
\begin{equation}
h_\beta = -\lim_{t\rightarrow\infty}\frac{1}{t}\frac{1}{\beta-1}\ln\sum_i^r p_i^\beta(t)
\,
\label{RenEn}
\end{equation}
is called R\'enyi  entropy. Equation~(\ref{PAhEq}) becomes more familiar when $\beta=1$,
and $\langle A\rangle$ is the positive Lyapunov exponent of the dynamics. In that case,
${\cal{P}}(1)$ is (minus) the escape rate $\gamma_0$, while 
\beq
 h_1 = -\lim_{t\rightarrow\infty}\frac{1}{t}\sum_i^r p_i(t)\ln\,p_i(t) 
\eeq 
is the  information entropy. One can then write~(\ref{PAhEq})  as~\cite{Gaspbook,KantzGrass85,EckRue85}
\begin{equation}
h_1 = -\gamma_0 +  \lambda 
\,,
\label{LyapTherm}
\end{equation}
that relates information entropy, escape rate, and Lyapunov exponent $\lambda$.

\section{Finite-time Thermodynamics}
\label{FiniteTime}
Let us attempt to extend the thermodynamic formalism to finite time, out of equilibrium. 
I shall begin with a finite-time topological pressure,
of the type
\beq
{\cal{P}}_t(\beta) = \frac{1}{t}  \ln \sum_j w_j(t)e^{-\beta\,A^t_j} 
\,,
\eeq
which, technically, may no longer be considered a free energy, since we are now out of equilibrium. 
The factors $w_j$ are included in the previous expression for the topological pressure, so that the Gibbs probabilities 
for every trajectory throughout the Markov partition are now weighed: 
\beq
p_j(t) = \frac{w_j(t)e^{-\beta A^t_j}}{\sum_i w_i(t)e^{-\beta A^t_i}}
\,.
\ee{FTGibbs}
At this stage, the previous expression is an Ansatz for the probability of  the integrated observable to be $A^t_j$ in the subsystem (i.e. symbolic sequence) labelled by $j$, surmised consistently with the 
tilted probability (large-deviations) formalism in a dynamics context~\cite{ChetrTouch13,ChetrTouch15}.
The meaning of Eq.~(\ref{FTGibbs}) will become clearer later on in this section, when the weights $w_j(t)$ are recognized as phase-space densities transported by the dynamics via the Perron-Frobenius/Koopman  operator. 

In what follows, I distinguish $\beta$, inverse temperature, or more simply variable of the characteristic (moment-generating, partition-) function
$Z_t(\beta)~=~\sum_j~w_j(t)e^{-\beta~A_j(t)}$, from $q$, order of the
R\'enyi entropy. Then, I use the so-generalized expression for the topological pressure
\beq
{\cal{P}}_t(q,\beta) = \frac{1}{t}  \ln \sum_j w^q_j(t)e^{-q\beta\,A_j^t} 
\,,
\eeq
while it is still understood that ${\cal{P}}_t(\beta):={\cal{P}}_t(q=1,\beta)$.
The R\'enyi information is the entropy~(\ref{RenEn}) for finite time: 
\begin{widetext}
\bea
\nonumber
I_t(q,\beta) = -\frac{1}{t}\frac{1}{q-1}\ln\sum_i^r p_i^q(t)  &=& 
\frac{1}{1-q}\frac{1}{t} \ln \sum_j \frac{w_j^qe^{-q\beta\,A_j^t}}{\left[\sum_i w_ie^{-\beta A_i^t}\right]^q} 
\\ 
&=& \frac{1}{1-q}\frac{1}{t} \ln \frac{e^{{\cal{P}}_t(q,\beta)t}}{Z^q_t(\beta)} = \frac{1}{1-q}\left[{\cal{P}}_t(q,\beta)- \frac{1}{t}\ln Z^q_t(\beta)\right]
\,.
\label{Renyi_therm}
\eea
The previous can be rewritten as
\begin{equation}
I_t(q,\beta) = \frac{1}{1-q}\left[{\cal{P}}_t(q,\beta) - q\,{\cal{P}}_t(\beta)\right]
\,,
\label{IPA}
\end{equation}
keeping in mind that everything is time dependent, and out of equilibrium. In the limit $q\rightarrow1$, we have that
\begin{equation}
I_t(1,\beta) = 
{\cal{P}}_t(\beta) - {\cal{P}}_t'(\beta) 
= \frac{1}{t}\left[ \ln  \sum_j w_je^{-\beta\,A_j^t} -\left.\frac{d}{dq}\ln \sum_j w_j^qe^{-q\beta\,A_j^t}\right|_{q=1}\right]
\,,
\label{I1outeq}
\end{equation}
where
\begin{equation}
\left.\frac{d}{dq}\ln \sum_j w_j^qe^{-q\beta\,A_j^t}\right|_{q=1}
= \frac{\sum_j w_je^{-\beta\,A_j^t}\left[\ln w_j - \beta A_j^t\right]}{Z_t(\beta)}  
\,.
\label{ddqlog}
\end{equation}
We can already spot three terms in the right-hand side of Eq.~\refeq{I1outeq}. The first is simply ${\cal{P}}_t(\beta)$ with the new definition of weighted 
probabilities for the symbolic sequences. The second is spelled out in Eq.~\refeq{ddqlog}:
\beq
\frac{1}{t}\frac{\sum_j \beta A_j^t w_je^{-\beta\,A_j^t}}{\sum_j w_je^{-\beta\,A_j^t}} =\frac{1}{t} \frac{\sum_j \beta p_j(t)A_j^t}{\sum_j p_j(t)} = \beta\langle A^t \rangle_\beta
\,,
\ee{FTthermAv} 
\end{widetext}
where the average is taken with respect to the definition~\refeq{FTGibbs} of the time-dependent probabilities. In that sense, the previous is a thermodynamic average. The third term
in the right-hand side of Eq.~\refeq{I1outeq} I also read out of Eq.~\refeq{ddqlog}:
\beq
\frac{\frac{1}{t}\sum_j w_j\ln w_j\,e^{-\beta\,A_j^t}}{\sum_j w_je^{-\beta\,A_j^t}} \equiv \frac{S_t(\beta)}{Z_t(\beta)}
\,.
\label{tiltinf}
\eeq
This quantity is peculiar of the finite-time thermodynamics, as it was not present in the equilibrium entropy-free'energy relation.
I name the numerator $S_t(\beta)$ of Eq.~\refeq{tiltinf} \textit{tilted information}. Now Eq.~\refeq{I1outeq} takes the form
\beq
I_t(1,\beta) = {\cal{P}}_t(\beta) + \beta\langle A^t \rangle_\beta - \frac{S_t(\beta)}{Z_t(\beta)}
\,.
\ee{I1outeqS}
Equation~\refeq{I1outeqS} is the first result of this paper: it generalizes the known asymptotic relation~\refeq{PAhEq} between information entropy, 
topological pressure, and expectation of an integrated observable to finite-time chaotic dynamics, out of statistical equilibrium. Besides the quantities mentioned,
the term $S_t$ newly appears in the equation: it represents the information provided by the weights $w_j$, biased by the Gibbs probabilities $e^{-\beta A_j^t}$ of the integrated
observable. This additional term must be transient, that is it must vanish as $t\rightarrow\infty$, when the whole expression~\refeq{I1outeqS} approaches the equilibrium 
identity~~\refeq{PAhEq}, as shown in Appendix~\ref{sec_asymp}.

At first sight, Eq.~\refeq{I1outeqS} does not bear physical significance beyond that of its equilibrium counterpart, nor
does it have ostensible implications or use. Yet, all the quantities involved are now defined in terms of the time-dependent statistical weights $w_j$,
that will play a central role in the non-equilibrium statistics of the integrated observables $A^t$, as shown in Sec.~\ref{AvSec}.

Let us now identify the weights $w_j$. This task is best accomplished by
taking as integrated observable the finite-time Lyapunov exponent
\beq
A^t = 
\ln ||J^t(x_0)|| = 
\ln |\Lambda(t)| 
\,,
\label{FTLE}
\eeq 
where the Jacobian is defined as
\begin{equation}
 J^t_{ij}(x) = \frac{\partial f_i^t(x)}{\partial x_j}
\,.
\label{Jac}    
\end{equation}
The regions of a given partition are evolved by the dynamics, 
in such a way that the total probability (measure) is conserved at every time step:
\beq
p_j(t) = \sum_i p_i(t+1)
\,.
\eeq 
Using our hypothesis~(\ref{FTGibbs}) for the $p_i$'s, the previous identity  
would translate to the evolution
 \begin{equation}
\frac{w_j(t)}{ |\Lambda_j(t)|^\beta} =  e^{-{\cal{P}}_t(\beta)} \sum_i \frac{w_i(t+1)}{{ |\Lambda_i(t+1)|^\beta}}
\,,
\label{prob_evol}
\end{equation}
once we write the probability of every partition element in terms of the stability exponent as integrated observable [Eq.~\refeq{FTLE}].
This evolution can be written in terms of the Perron-Frobenius operator~\cite{Slip13}, where we consider the $w_j$'s as densities. For a one-dimensional map, that reads
\begin{equation}
\rho'(x_j) = e^{-{\cal{P}}_t(\beta)} \sum_i \frac{\rho(x_i)}{|f'(x_i)|^\beta}
\,.
\end{equation}
That identifies the $w_j(t)$'s with time-varying densities.

Now, look at Eq.~\refeq{prob_evol}: going from time $t$ to time $t~+~1$ means to extend every interval of a Markov partition
by one symbol. For instance, if the symbolic dynamics is binary, 
that is the phase space is partitioned into two regions coded by $0$ and $1$, we let
$t=3$ and, say, $j=001$, then $~t~+~1~=~4$ and the terms of the summation on the right-hand side are
$i=0010, 0011$. In that case, the densities $w_j$'s map forward as stated by the `Perron-Frobenius' equation~\refeq{prob_evol}.
If we decide to stop at some definite $t$ and identify the set of all the distinct trajectories with a Markov partition, the $p_j$'s are given by the thermodynamic expression~\refeq{FTGibbs},
where the $w_j$'s are the densities in each interval of the partition. Therefore, the $w_j$'s are the 
probabilities of the dynamics to visit each region of the partition, or, equivalently, the probabilities of each trajectory,
whereas the $p_j$'s are the probabilities for region (sequence) $j$ to measure the observable $A_j^t$,
and, for finite time, they ought to be weighed by $w_j$, even if the dynamical system $f$ is ergodic
and mixing. 

Remarks are in order about the finite-time average and topological pressure. 
First, for $A^t~=~\ln~|\Lambda(t)|$, 
the finite-time thermodynamic average~\refeq{FTthermAv}
becomes
\beq
\langle \ln|\Lambda(t)| \rangle_\beta =  
\frac{\sum_j w_j(t)\ln\Lambda_j(t)/\Lambda_j^\beta(t)}{\sum_j w_j(t)/\Lambda_j^\beta(t)}
\,,
\label{LamAv}
\eeq
which  makes sense dynamically, if we recall that the neighborood of each region  
of a Markov partition of $t$ regions scales as $1/|\Lambda_j(t)|^d$ in a hyperbolic system~\cite{Gaspbook}. That way, Eq.~(\ref{LamAv}) may be
regarded as a weighed average over the partition.    
On the other hand, 
the finite-time topological pressure 
may be reconnected to more familiar quantitities such as the moment-generating function in a 
continuous phase space, where it is written as
\bea
\nonumber
{\cal{P}}_t(\beta) &=& \frac{1}{t}  \ln \sum_j w_j(t)e^{-\beta\,A_j^t}  \\ \nonumber
&\rightarrow&  \frac{1}{t}  \ln \int dx\,w(x(t)) e^{-\beta\,A^t(x)}  =  \frac{1}{t}  \ln \langle e^{-\beta A^t}\rangle 
\,, \\
\label{TopPcont}
\eea
where the first identity holds in the limit of an infinitely fine partition, with every point in the phase space
belonging to a distinct sequence of $t$ symbols, while 
the derivatives of $\langle e^{-\beta A^t}\rangle$
with respect to $\beta$ are the moments of the observable $A^t$ weighed by $w(x(t))$~\cite{dasBuch}.    
One may then rewrite Eq.~\refeq{I1outeqS} as
\beq
I_t(1,\beta) = \frac{1}{t}  \ln \langle e^{-\beta A}\rangle + \beta \langle A^t \rangle_\beta -  \frac{S_t(\beta)}{Z_t(\beta)}
\,,
\eeq
recalling that $I_1(\beta)$ and the last two terms do carry a prefactor of $1/t$ in their definitions.

Expressing the probability weights $w$ as phase-space densities in the continuum limit is central in the
evaluation of the finite-time averages of the integrated observables, as exposed in what follows.

\section{Averages}
\label{AvSec}
In this finite-time thermodynamic picture, an average is taken as in Eq.~\refeq{FTthermAv}, with the
numerator of that expression being
\beq
\sum_j  A^t_j w_j e^{-\beta A^t_j}
\,.
\label{Abackforth}
\eeq
The index $j$ refers to a particular orbit, identified with a symbolic sequence, and thus to 
a family of trajectories that share the same itinerary throughout the partition up to time $t$.
First, recall the definition~(\ref{E_intobs})
\begin{equation}
A^t(x)=  \sum_{\tau=0}^t a\left[f^\tau(x)\right]
\,.
\tag{\ref{E_intobs}}
\end{equation}
For a long sequence in a Markov partition, I shall approximate the sum over sequences with an
integral over the phase space, as seen in Eq.~(\ref{TopPcont}). In the
phase-space analog of Eq.~(\ref{Abackforth}), I can either 
choose $w$ to be computed at $t=0$, or at a later $t$. In the former case, the average may be written 
\beq
\int dx\, A^t(x) w(x) e^{-\beta A^t(x)}
\,,
\ee{av_numer}
whereas in the latter, the integrated observable follows the family of trajectories $f^{-t}(x)\rightarrow x$~\cite{footnote2},  
so that
 \beq
\int dx\, A^t(f^{-t}(x))e^{-\beta A^t(f^{-t}(x))}\left[ \Lop_t  w\right](x)  
\,,
\ee{av_fwd}
where  the Perron-Frobenius evolution operator acts as
\bea
\nonumber
 \left(\Lop_t \cdot w\right) (x) &=& \int dy \delta(x-f^t(y)) w(y) = \\  &=& \sum_{x_0=f^{-t}(x)}\frac{w(x_0)}{|\det J^t(x_0)|}
 \,.
 \eea
 \subsection{Going forward}
 Let us first study the average with the weight function evaluated at time $t$, as in Eq.~(\ref{av_fwd}), while Eq.~(\ref{av_numer}) with the density evaluated at time zero will be considered in
 Sec.~\ref{KoopPin}. Assuming a discrete spectrum for the Perron-Frobenius operator, the goal is now to
 express the average 
 \beq
 \langle A^t\rangle_\beta = \frac{\int dx\, A^t(f^{-t}(x))e^{-\beta A^t(f^{-t}(x))}\left[ \Lop_t  w\right](x)  }{\int dx\,e^{-\beta A^t(f^{-t}(x))}\left[ \Lop_t  w\right](x)  } 
 \label{ThermAv}
 \eeq 
 in terms of the leading eigenfunctions of $\Lop^t$.  
 \begin{widetext}
  In Eq.~\refeq{av_fwd} I now expand $\Lop_t w$, obtaining
   \beq
   \int dx\, A^t(f^{-t}(x))e^{-\beta A^t(f^{-t}(x))}\left[ \Lop_t  w\right](x)  = \sum_n^\infty b_n e^{-\gamma_n t} \int dx\, \phi_n(x)  A^t(f^{-t}(x)) e^{-\beta A^t(f^{-t}(x))}
   \,,
   \label{LopExpans}
   \eeq
   where $e^{-\gamma_n t}$ and $\phi_n(x)$ are respectively eigenvalues and eigenfunctions of the 
   Perron-Frobenius operator $\Lop_t$, while
   \beq
   b_n = \int w(y)\varphi_n(y) dy
    \ee{bn}
   are the coefficients of the expansion, that depend on the initial densities $w(x)$ and the eigenfunctions $\varphi_n$ of the Koopman operator $\Lop^\dagger_t$.
   Let us then go back to the original thermodynamic average~\refeq{ThermAv}, and rewrite it harnessing the above expansion:
   \beq
   \langle A^t \rangle_\beta = \frac{\sum_n^\infty b_n e^{-\gamma_n t} \int dx \, \phi_n(x)  A^t(f^{-t}(x)) e^{-\beta A^t(f^{-t}(x))}}{\sum_n^\infty b_n e^{-\gamma_n t} \int dx \, \phi_n(x) e^{-\beta A^t(f^{-t}(x))}}
 \,.
  \ee{Aexpansion} 
   Now let $\beta=0$:
      \beq
   \langle A^t \rangle_{\beta=0} = \frac{\sum_n^\infty b_n e^{-\gamma_n t} \int dx\, \phi_n(x)  A^t(f^{-t}(x))} {\sum_n^\infty b_n e^{-\gamma_n t} \int dx\, \phi_n(x)}
 \,.
  \eeq 
  It is the expansion of an average, and perhaps that does not say much per se.
  However, we may recover the  original expression for $A^t$ and focus on its phase-space dependence, if we start with an initial density of the type
  $w(y)~=~\delta(y~-~x_0)$, that is concentrated in one point:
   \beq
  \langle A^t \rangle_{\beta=0} (x_0) =
    \frac{ \int dx \, A^t(f^{-t}(x)) \int dy \, \delta(x-f^t(y)) \delta(y-x_0)}{ \int dx\, \int dy\, \delta(x-f^t(y)) \delta(y-x_0)} = \frac{A^t(x_0)}{\mu(\pS(t))} =: \hat{A}^t(x_0)
   \,,
   \label{obsdistA}
   \eeq
   where $\mu(\pS(t))$ is the fraction of trajectories that do not escape after time $t$, and it equals unity for a closed system. Concerning the expansion~\refeq{Aexpansion}, the coefficients $b_n$ defined by Eq.~\refeq{bn} simply equal $\varphi_n(x_0)$ when $w(x)=\delta(x-x_0)$, and the integrated observable spells
    \beq
     \langle A^t \rangle_{\beta=0}(x_0) = \hat{A}^t(x_0)  =   \frac{\sum_n^\infty \varphi_n(x_0) e^{-\gamma_n t} \int dx\, \phi_n(x)  A^t(f^{-t}(x))} {\sum_n^\infty \varphi_n(x_0) e^{-\gamma_n t} \int dx\, \phi_n(x)} 
    \,.
   \ee{Atexp}
In the limit $t\rightarrow\infty$, only the first term survives
     \beq
    \langle A^t \rangle_{\beta=0}(x_0) \rightarrow   \frac{ \int dx\,  \phi_0(x)  A^t(f^{-t}(x))} { \int dx\,  \phi_0(x)} 
    \,,
   \eeq
  that is simply the phase-space average weighed by the invariant density $\phi_0(x)$, as it is known at equilibrium, and the dependence on $x_0$ has been lost. 
  The interesting time scale in the present context is rather that of $(\gamma_1-\gamma_0)^{-1}$, at which the expanded average~\refeq{Atexp} is approximately
  \begin{equation}
 \hat{A}^t(x_0) :=  \langle A^t \rangle_{\beta=0}(x_0) 
\simeq  \int_\pS A^t(f^{-t}(x)) \phi_0(x) \,dx   +  \frac{\varphi_1(x_0)}{\varphi_0(x_0)}e^{-(\gamma_1-\gamma_0)t} \int_\pS A^t(f^{-t}(x)) \phi_1(x) \,dx\,.
\label{Atgamma1}
\end{equation}  
   \end{widetext}
  Now assume that the system has no escape. We may take the natural measure $\phi_0(x)$ to be $L^1$-normalized
  in the phase space, while the other eigenfunctions of the Perron-Frobenius operator are such that
  $\int dx\,\phi_n(x) = 0$.  That way, the previous expression for the `pointwise average' becomes
   \beq
     \hat{A}^t(x_0) \simeq   \langle A^t\rangle_{\pS} + \varphi_1(x_0) e^{-\gamma_1 t} \int dx\, \phi_1(x)  A^t(f^{-t}(x)) 
    \,,
   \ee{Atgamma1noesc}    
   where $ \langle A^t\rangle_{\pS} = \int dx \phi_0(x)  A^t(f^{-t}(x))$.
   In case of no escape ($\gamma_0=0$), the ground state of the Koopman operator is a uniform distribution, since 
   \beq
   \Lop^\dagger_t \varphi_0(x) = \varphi_0(f^t(x)) = \varphi_0(x)
   \eeq
   for every $x$, whence $\varphi_0(x)=1$ in Eq.~\refeq{Atgamma1noesc}.

Equation~\refeq{Atgamma1noesc} tells us that the phase-space profile of the integrated observable $A^t$ is entirely ruled by the subleading eigenfunction $\varphi_1(x)$ of the Koopman operator at the time scale $1/\gamma_1$, which determines $\hat{A}^t(x_0)$ independently of the observable itself. 

\subsection{Perron-Frobenius vs. Koopman operator}
The Perron-Frobenius operator $\Lop_t$ carries a density $\rho$, supported on $\pS$, forward in time 
to a density supported on a subset of $f^t(\pS)$~\cite{LasMac}. In this sense, $\Lop_t$ `follows the flow'.
One the other hand, the Koopman operator $\Lop_t^\dagger$ acts as 
\beq
\Lop_t^\dagger \rho(x) = \rho(f^t(x))
\,,
\eeq
and so
\beq
\rho(f^t(x)) = 0 \hspace{1cm} \mathrm{if}\, f^t(x) \notin \pS
\,.
\eeq
 That implies that
 \beq
\Lop^\dagger_t\rho(x) = 0 \hspace{1cm} \mathrm{if}\, x \notin f^{-t}(\pS)
\,,
\eeq
meaning that the the Koopman operator is supported on the preimage of the set $\pS$,
and thus $\Lop^\dagger_t$ may be thought of as transporting a density supported on $\pS$
backward in time to a density supported on $f^{-t}(\pS)$.  
\newline

\subsection{Going backward : a matter of pinning}
\label{KoopPin}
But why is the field profile of the integrated observable $A^t$ governed by the eigenfunctions of the   Koopman operator, and not by those of its adjoint, the Perron-Frobenius operator? The reason is that 
in Eq.~\refeq{av_fwd} $x$ is the final point of the density $w$ and of the 
integrated observable $A^t$. If, on the contrary, I had chosen to pinpoint density and observable by their values at the initial point of the trajectory originally labelled by the symbolic sequence $j$ in the discretized state space, I could have written $w$ as an evolution by the Koopman operator: 
 \beq
 w(f^t(x)) = \int dy \delta(y-f^t(x)) w(y)  
 \,.
 \eeq
Then, the course of action leading to the expansion of the average~\refeq{Aexpansion} is `adjointed'.
Let us use the definition~\refeq{E_intobs} for the integrated observable as a function of $x_0$, that is the initial point,
and rewrite Eq.~\refeq{av_numer} with observable and density still a function of the initial point $x$:
\begin{widetext}
\beq
  \int dx\, A^t(x) e^{-\beta A^t(x)} \left[  \Lop_t^\dagger w\right](x)
  = \int dx\,  e^{-\beta A^t(x)} A^t(x) \int dy \delta(y-f^t(x)) w(y)
  \,,
  \ee{ALop}
  \end{widetext}
 The density evolution in the left-hand side of~\refeq{ALop} is then expanded in terms of the eigenspectrum of the Koopman operator $\Lop_t^\dagger$,
 as done in Eq.~\refeq{LopExpans} for the Perron-Frobenius operator.
 As imaginable, the average of $A^t$ is now `dual' to the expression~\refeq{Aexpansion}, with the eigenfunctions $\varphi_j$ 
 of the Koopman operator replacing those of the Perron-Frobenius operator. Subsequently, I pin the initial density $w(x)=\delta(x-x_t)$
 at the arrival point of the trajectory $f^{-t}(x_t)\rightarrow x_t$, to finally obtain the quantity $\hat{A}^{-t}(x_t)$ (details in Appendix~\ref{BackEvol})
 in terms of the first two eigenfunctions $\phi_0$ and $\phi_1$ of the Perron-Frobenius operator:
 \begin{widetext}
 \begin{equation}
 \hat{A}^{-t}(x_t) :=
\frac{A^t(f^{-t} (x_t))}{\mu(\pS(t))|\det J^t(f^{-t}(x_t))|} \simeq  \int_\pS A^t(x) \varphi_0(x) \,dx \,   +
\frac{\phi_1(x_t)}{\phi_0(x_t)}e^{-(\gamma_1-\gamma_0)t} \int_\pS A^t(x) \varphi_1(x) \,dx\,.
\label{Aadjdistrunc}
\end{equation} 
\end{widetext}
The previous expression is akin to Eq.~\refeq{Atgamma1}, and it only but importantly differs in the 
variable of our choice (the arrival point of each phase-space trajectory, as opposed to the starting point in Eq.~\refeq{Atgamma1}),
as well as in the eigenfunctions of $\Lop_t$, rather than those of $\Lop^\dagger_t$.

 \subsection{Evolution on the manifold}
 An argument was presented in~\cite{LSYY21,YYSL21} to show that the evolution of a density by the Perron-Frobenius operator along the unstable manifold of an area-preserving, fully chaotic map leads to the 
following result: the distribution of the finite-time Lyapunov exponents in the phase space (the unit torus) follows the pattern of the second eigenfunction of the Perron-Frobenius operator, at a suitable time scale.
The idea is that, on the unstable manifold, the Perron-Frobenius operator acts on a density as
\beq
\Lop_t w(x) \sim e^{-t\Lambda(f^{-t}(x),t)} w(x)
\,,
\ee{Lop_um}   
where the Lyapunov trajectory begins at $f^{-t}(x)$ and runs for time $t$ up to $x$.
 On the other hand, as seen, the action of the Perron-Frobenius operator on the same density may be expanded and truncated as (assuming no escape)
 \beq
 \Lop_t w(x) = c_0 + c_1e^{-\gamma_1 t}\phi_1(x) + O\left(e^{-\gamma_2t}\right)
 \,.
 \eeq  
 Along the unstable manifold, the evolution is thus linearized as in Eq.~\refeq{Lop_um}, and so it does not depend on $w(x)$, but only on the finite-time Lyapunov exponent $e^{-\Lambda(f^{-t}(x),t)}$, whose phase-space profile should then follow $\phi_1(x)$ at a time scale set by $\gamma_1^{-1}$:
 \beq
  e^{-t\Lambda(f^{-t}(x),t)} \propto \phi_1(x)e^{-\gamma_1t}
 \,,
  \eeq
 meaning that the second eigenfunction of the Perron-Frobenius operator rules the distribution of the finite-time Lyapunov exponents pinned at the final point of each trajectory. That is consistent with Eq.~\refeq{Aadjdistrunc}, which generalizes the theory to an arbitrary observable (one would use Eq.~(\ref{Atgamma1}) when pinning the observable
 at the initial point of each iterated trajectory, instead). 
 
 \subsection{Noise}
 The finite-time thermodynamic formalism exposed in the previous section may also describe a
 chaotic system with background noise, according to the evolution
 \beq
 x_{t+1} = f(x_t) + \eta(t) := f_\eta(x_t)
 \,,
 \ee{Langev}
 with random force $\eta(t)$.  
 The integrated observable~(\ref{E_intobs}) would now take the form 
  \beq
 A_{\sigma^2}^t(x_0) = \left< \sum^t_{\tau=0} a(f^\tau_\eta (x_0)) \right>_{\sigma^2}
 \,,
 \ee{noisyA}    
where $\langle\cdot\rangle_{\sigma^2}$ denotes an ensemble average over noisy trajectories $f^t_\eta (x)$,
 with isotropic noise of amplitude $2\sigma^2$.  
 In this setting, phase-space
 densities are transported by an evolution operator with a noisy kernel, for 
 example the Fokker-Planck operator~\cite{CviLip12}, applied at each iteration:
 \beq
 [\Lop_{\sigma^2}\,w](x) = \frac{1}{\sqrt{4\pi \sigma^2}}\int dx\, e^{-(y-f(x))^2/4\sigma^2}w(x)
 \,.
 \label{FPevol}
 \eeq
\begin{widetext}
 With that change, Eq.~\refeq{Atexp} would become, in the noisy phase space,
 \beq
  \langle A^t \rangle_{\beta=0}(x_0) = \frac{A^t_{\sigma^2}(x_0)}{|\pS_{\sigma^2}(t)|} = 
   \frac{\sum_n^\infty \varphi_n(x_0) e^{-\gamma_n t} \int dx \phi_n(x)  A_{\sigma^2}^t(f^{-t}(x))} {\sum_n^\infty \varphi_n(x_0) e^{-\gamma_n t}} 
  \,.
  \ee{Ax0noise}
\end{widetext}
Now $\phi$ and $\varphi$ are respectively right and left eigenfunctions of the Fokker-Planck operator~\refeq{FPevol}, that retains the spectral gap of  the Perron-Frobenius operator, under the same assumptions as in the deterministic picture.    

\section{Validation}
 The above predictions are now tested on different models of chaos, namely the Bernoulli map, the perturbed cat map,  the noisy 
Hamiltonian H\'enon map, and the noisy Ikeda map. 
\subsection{Bernoulli map}
\label{BerSec}
It is defined as 
\beq
 f(x) = 2x\,\mathrm{mod}\,1 = \left\{
  \begin{array}{c} 
  2x \hspace{0.9cm} 0\leq x< \frac{1}{2}  \\   
  2x -1 \hspace{0.5cm} \frac{1}{2}\leq x< 1  
\,.
\end{array}
\right.
\label{BerMap}
\eeq 
This one-dimensional, non-invertible map, features chaos everywhere on the unit interval, no escape, and it has a constant Lyapunov exponent equal to $\ln 2$.

The spectra of both the Perron-Frobenius and the Koopman operators are discrete with $L^2$ as function space~\cite{Driebe}, and  available analytically. The Perron-Frobenius operator acts on a density
at each time step as
\beq 
\Lop\rho(x) = 	\frac{1}{2}\left[ \rho\left(\frac{x}{2}\right) +  \rho\left(\frac{x+1}{2}\right)\right]
\,,
\label{BerPF}
\eeq
and it has the Bernoulli polynomials 
\bea
\phi_0(x) &=& 1 \\
 \phi_1(x) &=& x - \frac{1}{2} \\
 \phi_2 (x) &=& x^2 -x + \frac{1}{6}  \label{BerPFEig} \\ \nonumber
   &\ldots&
\eea
as eigenfunctions of eigenvalues $\gamma_n=2^{-n}$. The one-dimensional nature of the phase space makes the forward action~\refeq{BerPF} all expanding, while the backward, Koopman operator~\cite{Fox}
 \beq 
\Lop^\dagger\rho(x) = 	\rho(2x)\Theta\left(\frac{1}{2}-x\right) +  \rho(2x-1)\Theta\left(x-\frac{1}{2}\right)
\,,
\label{BerKoop}
\eeq
is everywhere squeezing (here $\Theta$ is the Heaviside step function). Its leading eigenfunction $\varphi_0(x)=1$ is again uniform on the unit interval, while the rest of the spectrum is made of the generalized functions
\beq
\varphi_j(x) = \frac{(-1)^{j-1}}{j!}\left[\delta_-^{j-1}(x-1) -\delta_+^{j-1}(x)\right]
\,,  
\label{BerKoopEig}
\eeq
for $j\geq1$, that is combinations of  Dirac delta functions and their derivatives. This behavior is 
peculiar of one-dimensional chaotic maps. Due to the result~\refeq{BerKoopEig},  we may not apply the expansion~\refeq{Atgamma1}, that yields an integrated observable $\hat{A}^t(x)$ in terms of the 
first two eigenfunctions $\varphi_0$ and $\varphi_1$ of the Koopman operator,  since  the present theory assumes the $\varphi_i$'s to be smooth. Instead, we may test the validity of the 
expression~\refeq{Aadjdistrunc}, that relates the polynomial eigenfunctions~\refeq{BerPFEig} to the      
integrated observable $\hat{A}^{-t}(x)$, pinned by the final points of the orbits $f^{-t}(x)\rightarrow x$.

I shall now take two test observables, $a_1(x)=x^2$, and $a_2(x)=x+\sin20x$, apply the Perron-Frobenius operator~\refeq{BerPF} and integrate the outcomes over a finite time interval.   
\begin{figure}[tbh!]
\centerline{
(a)\scalebox{.33}{\includegraphics{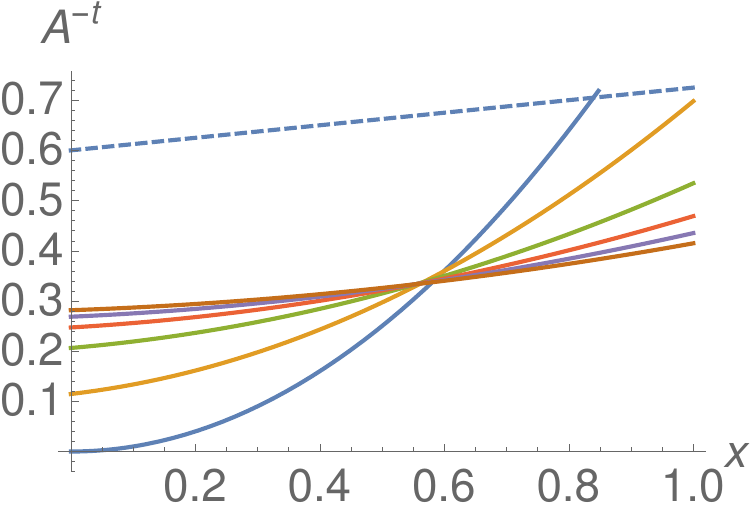}}
(b)\scalebox{.33}{\includegraphics{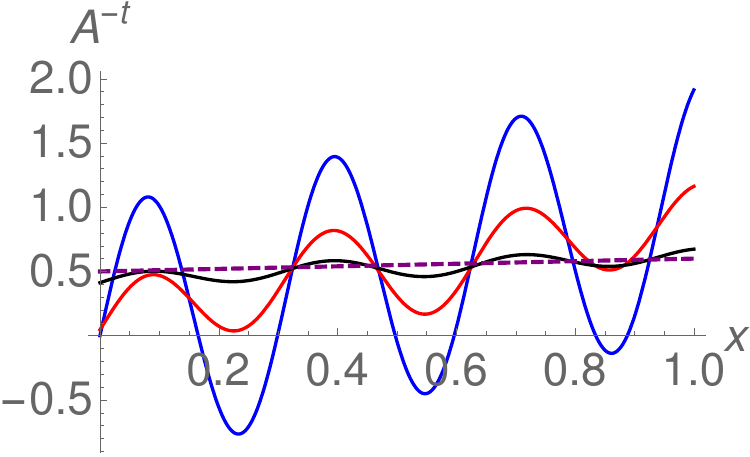}}}
\caption{Test observables mapped by the Perron-Frobenius operator and time integrated: (a) $a_1=x^2$ (blue, solid line) and successive $\hat{A}^{-t}_1(x)$ (solid lines in color)
with $t=3$, $t=5$, $t=8$, $t=10$, $t=15$ approaching a straight line (for increasing $t$), plotted above (dashed line) for comparison; (b) $a_2 = x +\sin20x$ (blue, solid line), and successive $\hat{A}^{-t}_2(x)$ (solid lines in color) with $t=8$, $t=15$ approaching a straight line (dashed line).}
\label{BerObs}    
 \end{figure}
Applying $\Lop_t$ has the effect of progressively smoothing observables as $t$ increases, e.g.  
\beq 
[\Lop_{t=5} a_1](x) = 0.3 + 0.06x+ 0.004x^2
\,,
\eeq  
or
\beq 
[\Lop_{t=5} a_2](x) = 0.46 + 0.14x+ 0.002x^2 + O(x^3)
\,,
\eeq  
when expanding the closed-form expression of $[\Lop_{t=5} a_2](x)$ in a power series. As a consequence, the mapping of both $a_1$, $a_2$ produces curves that
are approximately straight, and, at a longer timescale, the corresponding integrated observables $\hat{A}_1^{-t}$, $\hat{A}_2^{-t}$, also approach the functional form of the second eigenfunction of the 
Perron-Frobenius spectrum (Fig.~\ref{BerObs}),  $\phi_1(x) = x - \frac{1}{2}$. Specifically, Eq.~\refeq{Aadjdistrunc} predicts that $\hat{A}^{-t}(x) \sim \phi_1(x)/\phi_0(x)$, where $\phi_0(x)=1$ in 
this case, while  the proportionality factor (slope of the line) depends on the observable.     
\subsection{Two-dimensional maps}

The strategy is that to numerically compute the first two eigenfunctions of the Perron-Frobenius and of the Koopman spectrum, and compare their ratio with the phase-space profiles of different integrated
observables for finite time.
 
The leading- and subleading eigenfunctions of the Perron-Frobenius/Koopman operator 
are first computed as follows. The transfer operator is
projected onto a finite-dimensional vector space, and thus
implemented as a matrix, as it is by now common when 
solving flow (e.~g. Liouville~\cite{ChapTan13}) equations. Previous literature warns us
that the choice of the discretization is crucial~\cite{Froy07}, and may deeply
affect the eigenspectrum beyond the leading eigenvalue~\cite{Brini}. It has been established, on the other hand,
that nonlinear perturbations to linear maps on a torus increase the robustness of the numerically
evaluated spectrum under certain conditions~\cite{BKL02}.
The simplest discretization scheme is Ulam's method~\cite{Ulam},
that amounts to subdividing the phase space into $N$ intervals
$\pS_i$ of equal area. The evolution operator is thus approximated
with a $N\times N$ transfer matrix whose entries $\mathbf{L}_{ij}$ are the transition probabilities from 
$\pS_i$ to $\pS_j$
\beq
\mathbf{L}_{ij} = \frac{\mu\left(\pS_i \bigcap f^t(\pS_j)\right)}{\mu(\pS_i)}
\label{Ulmat}
\eeq
in one time step, where $\mu$ is the Lebesgue measure. I use a known
Monte Carlo method~\cite{ErmShep} to estimate the nonsymmetric transfer matrix $\mathbf{L}_{ij}$,
that consists of iterating random initial conditions from each cell $\pS_i$  and counting which fraction lands in each $\pS_j$.   
A thorough study of stability and convergence of
discretization algorithms has been reported elsewhere by the author and co-workers~\cite{YYSL21}, among
others.
 
\subsubsection{Perturbed cat map}
 \label{CAT}
The first two-dimensional model considered is the perturbed cat map $f(x)~=~T_\epsilon~\circ~T[x]$, with $x=(q,p)$:
 \beq
 T  \left( \begin{array}{c}
 q \\
 p
 \end{array} 
 \right)= \left(\begin{array}{cc}
 1 & 1 \\
 1 & 2 
 \end{array}
 \right)
  \left( \begin{array}{c}
 q \\
 p 
 \end{array}
 \right) \hskip0.2cm
 \mathrm{mod}\, 1
 \,,
 \eeq
 and
 \beq
 T_\epsilon 
 \left( \begin{array}{c}
 q \\
 p 
 \end{array}
 \right)
 = \left(\begin{array}{c}
 q - \epsilon\sin2\pi p \\
 p
 \end{array}
 \right) \hskip0.2cm
 \mathrm{mod} \,1
 \,.
 \label{CatNL}
 \eeq
 \begin{figure}[tbh!]
\centerline{
(a)\scalebox{.35}{\includegraphics{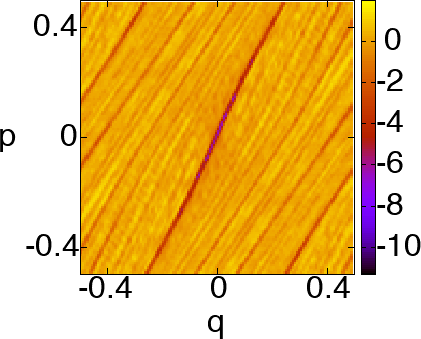}}
\hskip 0.1cm
(b)\scalebox{.35}{\includegraphics{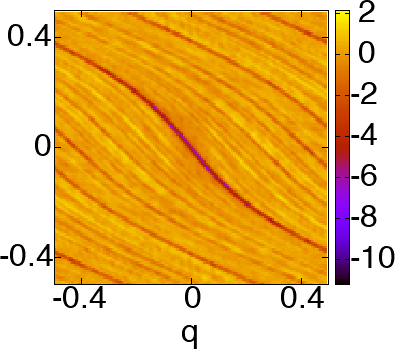}}}
\caption{First subleading eigenfunctions of the (a) Perron-Frobenius- and (b) Koopman operator for the perturbed cat map. The Ulam matrix has size $2^{14}\times2^{14}$.}
\label{Sublead}
\end{figure}
This system is strongly chaotic and hyperbolic, that is, correlations decay exponentially fast with time~\cite{Arnold}. It possesses an infinite number of unstable periodic orbits, 
and, specifically, a fixed point at the origin. The phase space is a 2-torus, there is no escape, and areas are preserved by the time evolution, so that the determinant of the Jacobian 
matrix of every trajectory is equal to unity. I set the parameter $\epsilon=0.1$ in what follows, a small enough value for the topology of the phase space to be 
preserved, yet large enough for the nonlinearity~\refeq{CatNL} to make the Ulam discretization robust.

The leading eigenfunctions of the transfer operators for the cat map are uniform distributions, and 
therefore they will not affect the predictions of the present theory out of equilibrium. Instead, the   
subleading eigenfunctions of the Perron-Frobenius and Koopman operators, computed with the Ulam
method, are shown in Fig.~\ref{Sublead}.

\begin{figure*}[tbh!]
(a)\scalebox{.4}{\includegraphics{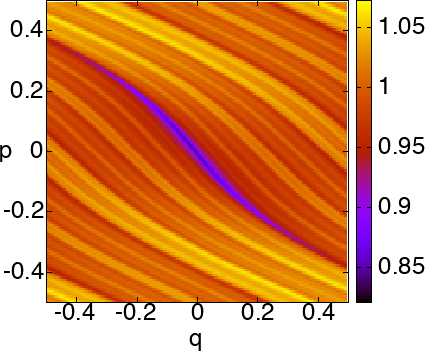}}
(b)\scalebox{.4}{\includegraphics{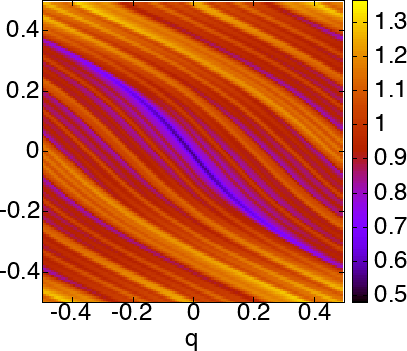}}
(c)\scalebox{.42}{\includegraphics{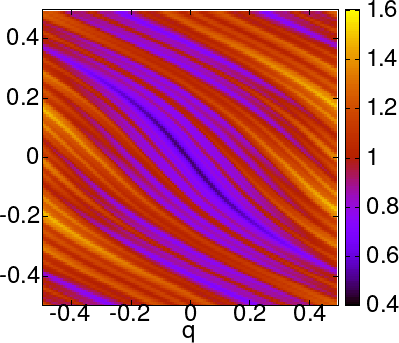}}
\caption{Phase-space density plots  ($2^{14}$ points, each averaged over $10^4$ trajectories)  of (a) the finite-time Lyapunov exponents, (b) the integrated kinetic energy, (c) the average diffusion, of the perturbed cat map having the
$(q,p)$ coordinates as initial points. The map is iterated until time $t=15$.}
\label{CatInitt5}
\end{figure*}

 Here, I first verify prediction~\refeq{Atgamma1noesc} for an integrated observable pinned by its initial point in the phase space, $A^t(x_0)$. The observables employed here are 
 \begin{enumerate} 
 \item The finite-time Lyapunov exponent~(\ref{FTLE}) of the cat map.
 \item The average diffusion
 \beq
 \overline{D}^t(x) = \frac{1}{t}\sum_{\tau=0}^t  q^2(f^\tau(x))
 \,,
 \eeq    
 with $t\sim\gamma_1^{-1}$.
 \item The average kinetic energy
 \beq
 \overline{K}^t(x) = \frac{1}{t}\sum_{\tau=0}^t \frac{p^2(f^\tau(x))}{2}
 \,,
 \eeq  
 again with  $t\sim\gamma_1^{-1}$.
 \end{enumerate}
 The desired finite-time density plots are obtained by iterating some $10^8$ randomly chosen, uniformly distributed initial conditions until a time $t$ before relaxation.  
 The outcomes are shown in Fig.~\ref{CatInitt5}: all the observables are striated along the stable manifold, and, in particular, their profiles are all alike, and their features echo with those of the first subleading eigenfunction of the Koopman operator [Fig.~\ref{Sublead}(b)]. Enhancement (`scar') of the latter~\cite{LSYY21} corresponds to suppression (`antiscar') of the integrated observable, which 
 is ascribed to the second term of Eq.~\refeq{Atgamma1noesc}: it is found that either $\varphi_1(x_0)<0$ and maximally negative at the scar with the integral $ \int dx \, \phi_1(x)  A^t(f^{-t})(x)) >0$ (from numerics), or vice versa~\cite{footnote3}, and so the pointwise value of the integrated observable $\hat{A}^t$ is approximately given by a constant (its phase-space average) \textit{minus} something proportional to the second eigenfunction of the Koopman operator.    As a result,
a scar in the second eigenfunction produces an antiscar in the 
 density plots of  all integrated observables at time scale $\gamma_1^{-1}$, as apparent in Fig.~\ref{CatInitt5}. 
  \begin{figure*}[tbh!]
(a)\scalebox{.4}{\includegraphics{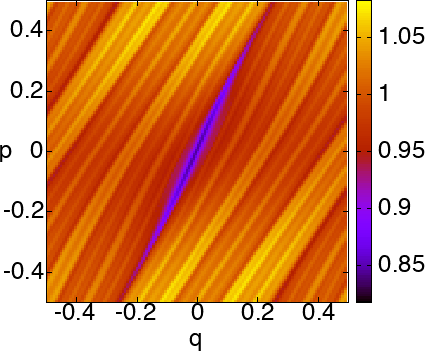}}
(b)\scalebox{.4}{\includegraphics{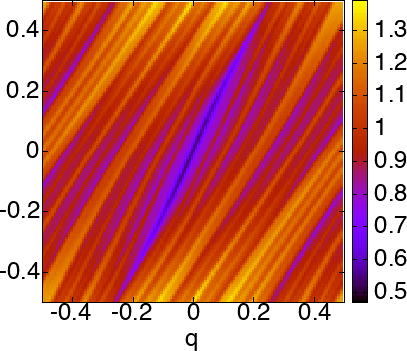}}
(c)\scalebox{.42}{\includegraphics{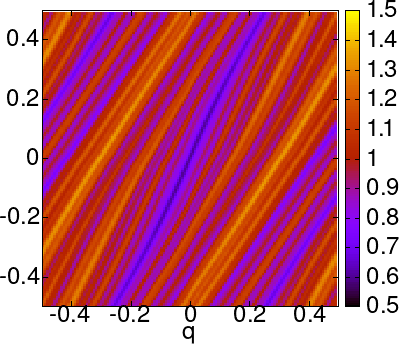}}
\caption{Same as Fig.~\ref{CatInitt5}, but here $(q, p)$ are the coordinates of the *final* points.
}
\label{CatFint5}
\end{figure*}

 Let me now proceed specularly with the validation of the prediction~\refeq{Aadjdistrunc} of an integrated observable pinned by its final point (iteration at time $t$) in the phase space, $\hat{A}^{-t}(x_t)$. The tested observables are once again the finite-time Lyapunov exponent, the average diffusion, and the kinetic energy.
 The results are displayed in Fig.~\ref{CatFint5}, and are analogous to what seen with the initial-point
 pinning, except that all the profiles are striated along the unstable manifold of the map, and follow the second eigenfunction of the Perron-Frobenius operator [Fig.~\ref{Sublead}(a)], here denoted by $\phi_1(x)$. 
 The equivalence between eigenfunction enhancement and observable suppression is still verified in
 this case,  as one can infer from Eq.~(\ref{Aadjdistrunc}).
     
\begin{figure}[tbh!]
\centerline{
(a)\scalebox{.38}{\includegraphics{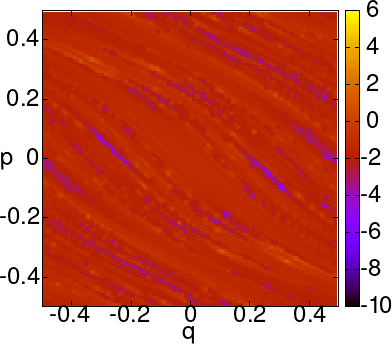}}
(b)\scalebox{.38}{\includegraphics{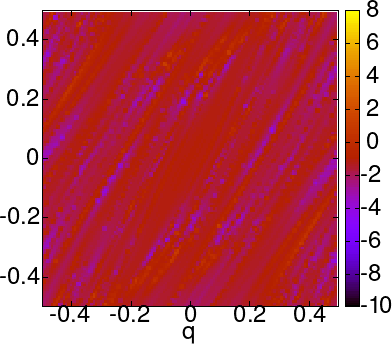}}}
\caption{The logarithmic ratio~\refeq{ratio} between ($2^{14}$ points) the finite-time Lyapunov exponents and the integrated kinetic energy, for the perturbed cat map having the
$(q,p)$ coordinates as (a) initial- and (b) final points. The map is iterated until time $t=15$.}
\label{CatRat}    
 \end{figure}

 In order to quantify the similarities between distinct finite-time
 observables, Fig.~\ref{CatRat} portrays the density plot of the logarithmic ratio 
 \beq
 r(x_0,t) = \ln \left|\frac{\ln|\Lambda(x_0,t)| - \langle\ln|\lambda(x,t)|\rangle_\pS}
 {\overline{D}(x_0,t) -\langle \overline{D}(x,t)|\rangle_\pS}\right|
 \ee{ratio}
 between the diffusion and the Lyapunov exponent, scaled by their mean values,
 both in the Perron-Frobenius and in the Koopman pictures of pinpointing. If the present theory holds, we should expect $r(x,t)$ to be a uniform distribution plus or minus fluctuations, and indeed the plots 
 give that indication.
  
 \subsubsection{Hamiltonian H\'enon map}
 \label{HamHen}
The next model to test the theory on is the Hamiltonian H\'enon map
 \bea
 \nonumber
 q' &=& 1 -\alpha q^2 +\beta p \\   
p'  &=& q
\,,
\label{HenMap}
\eea
  \begin{figure}[tbh!]
\centerline{
(a)\scalebox{.32}{\includegraphics{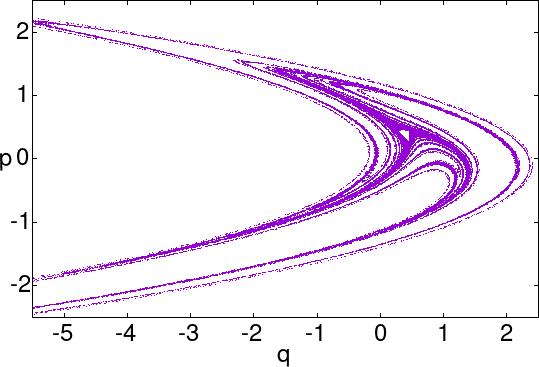}}
(b)\scalebox{.26}{\includegraphics{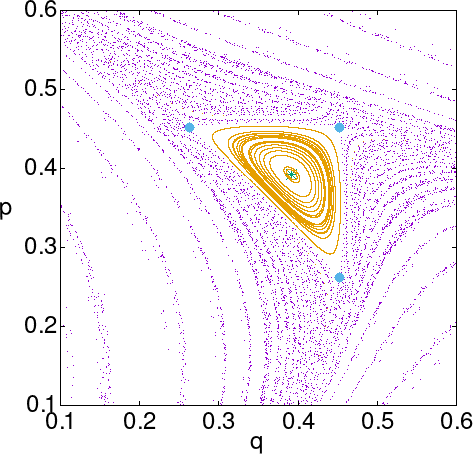}}}
\caption{(a) Unstable manifold emanating from the hyperbolic fixed point of the map~(\ref{HenMap}), obtained by forward iteration of $10^{6}$ initial conditions until time $t=15$.
(b) Marginally stable fixed point, surrounded by a stability island, triangularly shaped by an outer period-three unstable periodic orbit (blue points).}
\label{HenPhase}    
 \end{figure}
 with $\alpha=1.4$ and $\beta=-1$. This choice of the parameters avoids 
 dissipation in the dynamics ($|\beta|=1$), but it offers a new scenario to validate the present theory, due to: 
  $i)$ escape to infinity from the 
 neighborhood of the hyperbolic fixed point $x_p\simeq(-1.1,-1.89)$, that generates a chaotic saddle through
 its stable and unstable manifolds (the latter is portrayed in Fig.~\ref{HenPhase}(a)); $ii)$ a not-everywhere chaotic but rather mixed phase space, given the
 presence of   
 a second fixed point, $x_c\simeq(0.39,0.39)$,  that is marginally stable, and surrounded by a small non-hyperbolic region~(Fig.~\ref{HenPhase}(b)).
  In order to `kick' the dynamics out of the latter non-chaotic region and into the chaotic phase, weak noise is 
 added to the map~(\ref{HenMap}) of an amplitude comparable to the size of the stability island per unit time.
 
\begin{figure*}[tbh!]
\centerline{
(a)\scalebox{.5}{\includegraphics{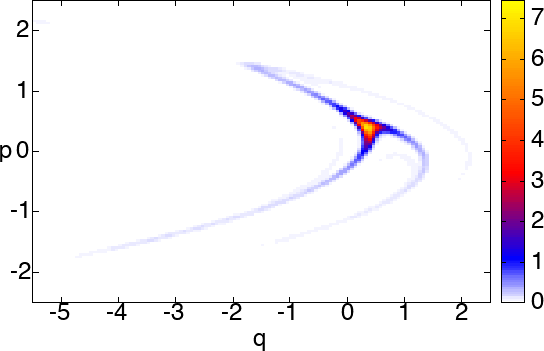}}
\hskip 0.7cm
(b)\scalebox{.5}{\includegraphics{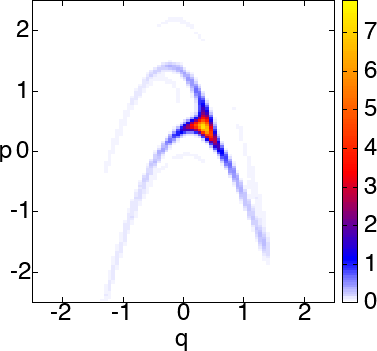}}}
\vskip 0.2cm
\centerline{
(c)\scalebox{.5}{\includegraphics{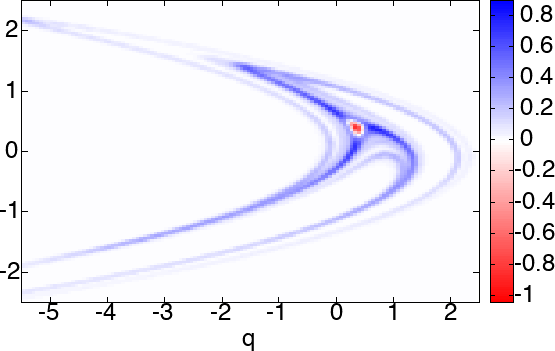}}
\hskip 0.7cm
(d)\scalebox{.5}{\includegraphics{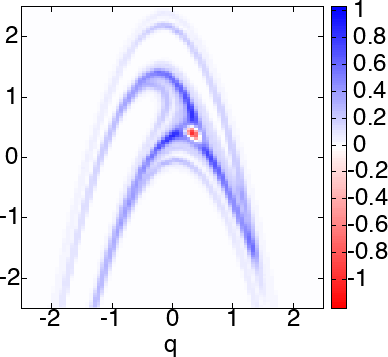}}}
\caption{(a-b): Leading eigenfunctions of the (a) Perron-Frobenius- and (b) Koopman operator for the H\'enon map.(c-d): First subleading eigenfunctions of the same operators, respectively.  
 The Ulam matrix has size $2^{14}\times2^{14}$.}
\label{heneigs}
\end{figure*}

 Strictly speaking, the non-hyperbolicity of the resulting noisy system should introduce a continuous component
 in the spectrum of the transport operators and thus break the assumption of a solely discrete spectrum. 
   However, if we investigate
 timescales of the order of- or shorter than the inverse escape rate from the region of the chaotic saddle, when the 
 discrete part of the spectrum is dominant, the contribution of the continuous part of the spectrum
 may be ignored, due to the smallness of the stability island.
 Unlike for the cat map, the leading eigenfunctions of the transfer operators for the Hamiltonian H\'enon map
 are not uniform distributions~(Fig.~\ref{heneigs}(a-b)), and, due to the finite escape rate, they are conditionally invariant densities.
 As a consequence, we should expect from the predictions~\refeq{Atgamma1} and~(\ref{Aadjdistrunc}) that the nonequilibrium profiles  of integrated observables 
 follow the ratio of the first subleading- to the leading eigenfuntion of $\Lop_t$ 
 ($\Lop^\dagger_t$).    

\begin{figure*}[tbh!]
\centerline{
(a)\scalebox{.5}{\includegraphics{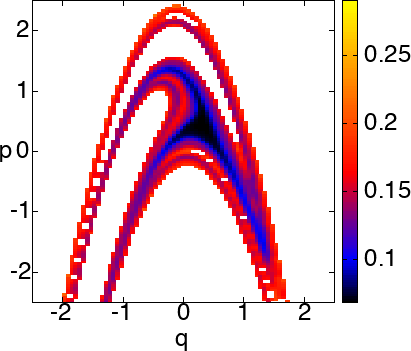}}
\hskip 0.7cm
(b)\scalebox{.5}{\includegraphics{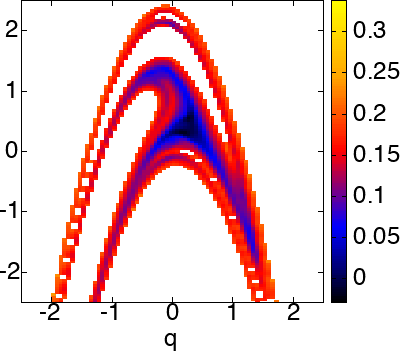}}}
\vskip 0.2cm
\centerline{
(c)\scalebox{.5}{\includegraphics{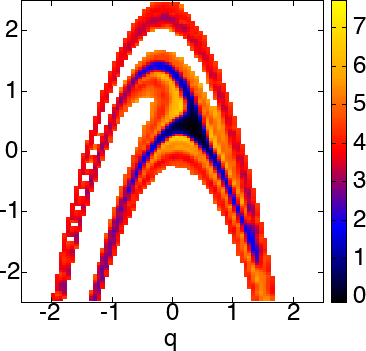}}
\hskip 0.7cm
(d)\scalebox{.52}{\includegraphics{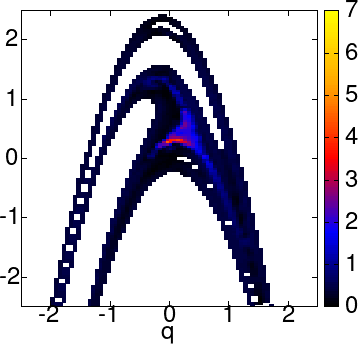}}}
\caption{(a): Phase-space density plot  ($2^{14}$ points, each averaged over $10^4$ trajectories)  of the diffusivity~\refeq{xyDiff} for the H\'enon map, pinned by the initial points, after $t=10$ iterations of the map. (b): The finite-time Lyapunov exponents, $t=10$. (c) Ratio of the first subleading-  to the leading eigenfunction of the Koopman operator for the same map.
(d) ratio of (a) to (b), as defined in~\refeq{ratio}.}
\label{henobs}
\end{figure*}

 The theory is tested for two observables, that is the finite-time Lyapunov exponent, as well as the diffusivity 
 \beq
 \hat{D}^t(x) = \frac{1}{t} \sum_{\tau=0}^t \, [q^2(f^\tau(x)) + p^2(f^\tau(x))]
 \,. 
\ee{xyDiff}
The density plots in Fig.~\ref{henobs} corroborate the expectations for the two integrated observables to be supported on the 
stable manifold of the map when pinned by the initial points of the iteration $x_0\rightarrow f^t(x_0)$ plus weak noise, and to mimic the ratio between the 
second and the first eigenfunction of the Koopman operator. 

\begin{figure*}[tbh!]
\centerline{
(a)\scalebox{.45}{\includegraphics{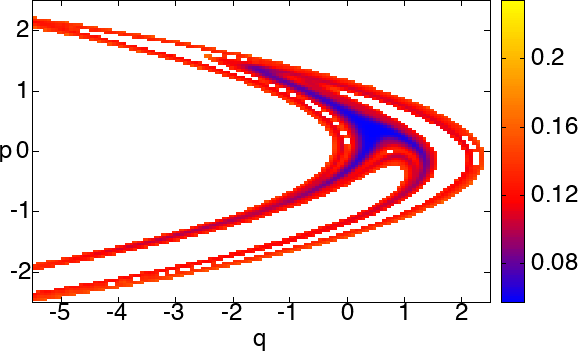}}
\hskip 0.3cm
(b)\scalebox{.45}{\includegraphics{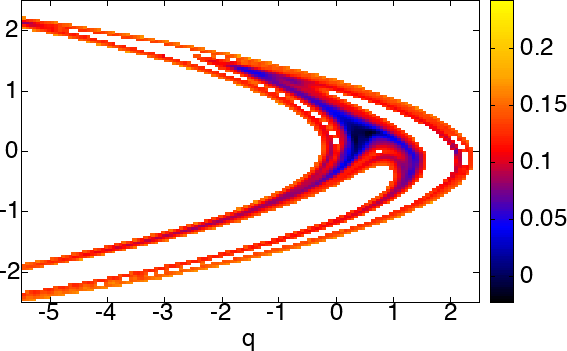}}}
\vskip 0.1cm
\centerline{
(c)\scalebox{.45}{\includegraphics{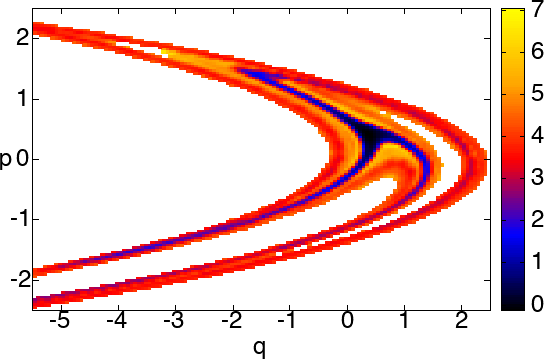}}
\hskip 0.3cm
(d)\scalebox{.47}{\includegraphics{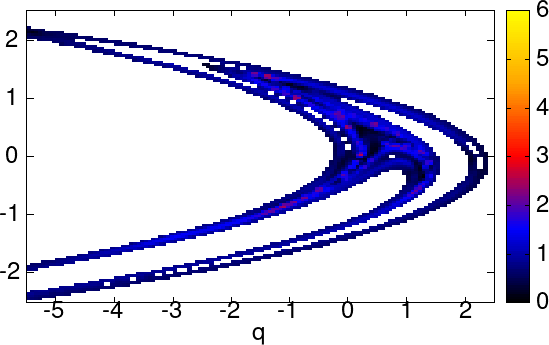}}}
\caption{(a): Phase-space density plot  ($2^{14}$ points, each averaged over $10^4$ trajectories)  of the diffusivity~\refeq{xyDiff} for the H\'enon map, pinned by the final points, after $t=10$ iterations of the map. (b): Distribution of the finite-time Lyapunov exponents, $t=10$. (c) Ratio of the first subleading-  to the leading eigenfunction of the Perron-Frobenius operator for the same map.
(d) ratio of (a) to (b),  as defined in~\refeq{ratio}.
}
\label{henbobs}
\end{figure*}

On the other hand, the same observables pinned by the final points 
of each phase-space trajectory $f^{-t}(x_t)\rightarrow x_t$ plus weak noise are supported on the unstable manifold of the map (Fig.~\ref{henbobs}), and behave similarly to the ratio of the second to the first eigenfunction of the Perron-Frobenius operator. In both `forward' and `backward' pictures, the strongly chaotic phase (in orange) is distinguishable from the non-hyperbolic, weakly chaotic phase (in blue) of a three-lobed shape with tapered ends, due to a period-three unstable periodic orbit that rules the dynamics just outside the 
stability island~(Fig.~\ref{HenPhase}(b)). 
   
Figures~\ref{henobs}(a)-(b) and~\ref{henbobs}(a)-(b) show that the phase-space profiles of the two observables 
 are nearly identical, and their ratios (Figs.~\ref{henobs}(d) and~\ref{henbobs}(d)) are
 very close to be uniform, except for deviations visible in the non-hyperbolic region.  

The white color in Figs.~\ref{henobs}-\ref{henbobs}(a)-(b) indicates the region of the phase space where forward (Fig.~\ref{henobs}) or backward (Fig.~\ref{henbobs}) trajectories escape from the domain examined before the time $t$ of integration. In Figs.~\ref{henobs}-\ref{henbobs}(c), instead, the ratio between the eigenfunctions is not defined in the blank region, where the leading eigenfunction vanishes. 

The density plots of the leading- and subleading eigenfunctions of the transport operators (Fig.~\ref{heneigs}) taken separately, bear significant differences from those of the observables: the first eigenfunctions clearly describe a longer timescale than that of the observables profiles, at which noisy trajectories have mostly left the hyperbolic region, while they only survive in and around the stability island; the second eigenfunctions alone are more resemblant of the finite-time integrated observables, except they are suppressed on a ring around the stability island. That pattern is not detected in the density plots 
of the observables. Therefore, it does appear as though the latter are 
best described by the ratio $\phi_1/\phi_0$ ($\varphi_1/\varphi_0$).                

\subsubsection{Ikeda map}
\label{SecIkeda}
Let me now consider the Ikeda map
\bea
\nonumber
q' &=& c_0 + c_2q\cos\theta - c_2p\sin\theta \\
p' &=& c_2q\sin\theta + c_2p\cos\theta 
\,,
\label{Ikeda}
\eea
with $\theta = c_1 - \frac{c_3}{1+q^2+p^2}$, while the parameters are set to
$c_0=1$, $c_1=0.4$, $c_2=0.9$, $c_3=6$. The Ikeda map with these parameters 
features a strange attractor, that is the closure of the unstable manifold of the fixed point 
at $x_s\simeq(0.5228,0.2469)$ (inverse saddle). The basin of attraction of the strange attractor is bounded by the
stable and unstable manifolds of the hyperbolic fixed point  $x_h~\simeq~(1.1142,-2.2857)$~\cite{Osipenko}.
 \begin{figure}[tbh!]
\centerline{
(a)\scalebox{.5}{\includegraphics{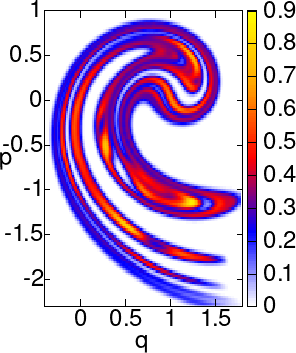}}
\hskip 0.1cm
(b)\scalebox{.5}{\includegraphics{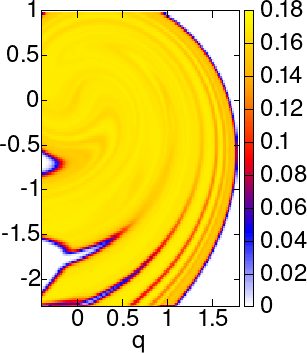}}}
\caption{First eigenfunctions of the (a) Perron-Frobenius- and (b) Koopman operator for the Ikeda map with additive noise of amplitude $\sigma^2=2.5\cdot 10^{-3}$. The Ulam matrix has size $2^{14}\times2^{14}$.}
\label{ikzero}
\end{figure}

The Ikeda map is dissipative, and thus it does not preserve areas in the phase space, which are instead
shrunk by the evolution. This phenomenon is known as squeezing, and it adds to the stretching and folding already seen in the previous Hamiltonian examples, but now plays a central role.

The first consequence of squeezing would be a strange attractor with a non-smooth measure, that emerges at long times.
In order to avoid fractal measures in the phase space, that would prevent the present theory from applying, Gaussian, uncorrelated, and isotropic
noise of amplitude  $2\sigma^2=0.025$ is added to the 
dynamics~(\ref{Ikeda}). Collaterally, some noisy orbits may now cross the stable manifold of the fixed point
 $x_h$, and exit the region of the strange attractor, which produces a tiny but non-zero escape rate.
 
\begin{figure*}[tbh!]
\centerline{
(a)\scalebox{.6}{\includegraphics{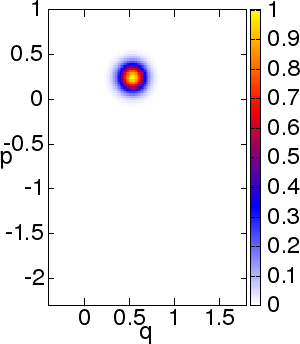}}
\hskip 0.1cm
(b)\scalebox{.6}{\includegraphics{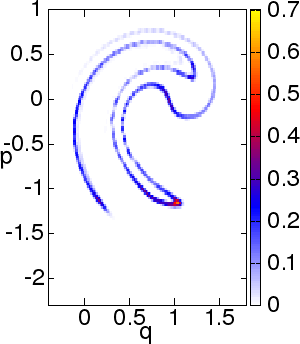}}
\hskip 0.1cm
(c)\scalebox{.6}{\includegraphics{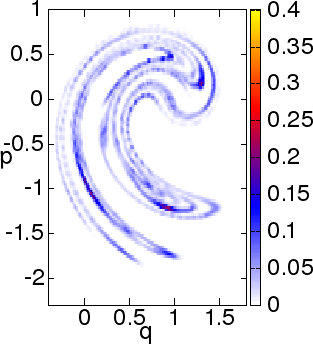}}}
\vskip 0.2cm
\centerline{
(d)\scalebox{.6}{\includegraphics{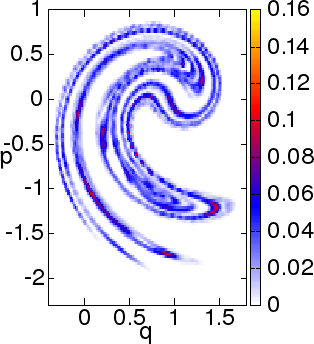}}
\hskip 0.1 cm
(e) \scalebox{.6}{\includegraphics{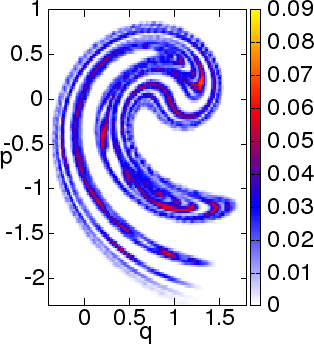}}
\hskip 0.1 cm
(f) \scalebox{.4}{\includegraphics{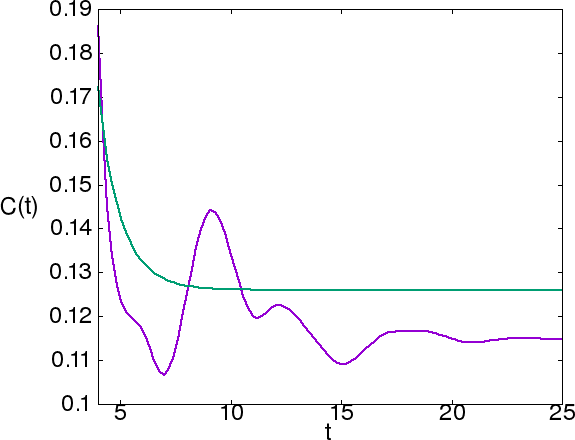}}
}
\caption{Snapshots of the evolution of the initial Gaussian density centered at the fixed point $x_s\simeq(0.533,0.247)$, portayed in (a), by the Ulam matrix of the 
Ikeda map with noise of amplitude $\sigma^2=4\cdot 10^{-4}$, given by the resolution; (b) $t=5$; (c) $t=9$; (d) $t=11$; (e) $t=25$. (f) (purple) autocorrelation function of the initial density
in (a) with the iterates at an intermediate timescale; (green) autocorrelation function of a density initially centered at the fixed point of the perturbed cat map, for comparison.   
 The Ulam matrix has size $2^{12}\times2^{12}$.}
\label{ikcorr}
\end{figure*}

 The second effect of squeezing is a complex-conjugate pair of second eigenvalues for the transport operators, instead of the real and isolated single eigenvalue encountered in the previous Hamiltonian models.
 The second eigenvalue of the Perron-Frobenius operator yields the decay rate of any initial density to
 the natural measure of the phase space, which can be estimated from the autocorrelation function 
\beq
C(t) = \frac{\int w(x) \left[\Lop_t w\right](x)}{\int w^2(x)}
\,.
\label{autocorr}
\eeq
 Here $C(t)$ is computed
 for an initial Gaussian density centered at the fixed point of the map, transported by the Ulam matrix $\mathbf{L}^t$,  which approximates $\Lop_t$,
 and plotted as a function of time in Fig.~\ref{ikcorr}(f), where it can be clearly seen to oscillate while decaying. 
 A non-trivial imaginary part of the first subleading eigenvalue of the Perron-Frobenius/Koopman operator signals oscillations in the decay of correlations, produced by the alternate effects of stretching, folding, and 
 especially squeezing (Figs.~\ref{ikcorr}(a)-(e)) that creates accumulation regions, unlike in the previous 
 Hamiltonian models,  
 where correlations decay monotonically (Fig.~\ref{ikcorr}(f)).  
 
 With that observation, let us look once more at 
 the first non-trivial term in the 
expansion~(\ref{Atgamma1}) of the integrated observable $A^t(x_0)$:
\beq
\varphi_1(x_0) e^{-\gamma_1 t} \int dx\, \phi_1(x)  A^t(f^{-t}(x))
\eeq
is now made of three complex factors, and adds up to a real number with the mirror term
of $\varphi_1^*(x_0)$ and $\phi_1^*(x_0)$. In that process, the combination of $\varphi_1(x_0)$
and $\varphi_1^*(x_0)$ gets to depend on  $\int dx\, \phi_1(x)  A^t(f^{-t}(x))$ and its complex conjugate,
which are observable specific. That in fact determines the breakdown of the universal behavior of the finite-time integrated observables predicted by the 
theory and exemplified in the previous models. For the Ikeda map, one should not expect
distinct observables to share the same profile, as they are supposed to follow 
different linear combinations of the real and imaginary parts of the first subleading eigenfunctions
(divided by the natural measure), all observable dependent.
 
\begin{figure*}[tbh!]
\centerline{
(a)\scalebox{.65}{\includegraphics{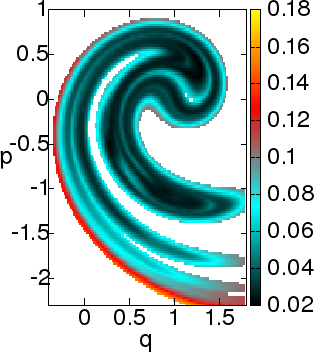}}
\hskip 0.3cm
(b)\scalebox{.65}{\includegraphics{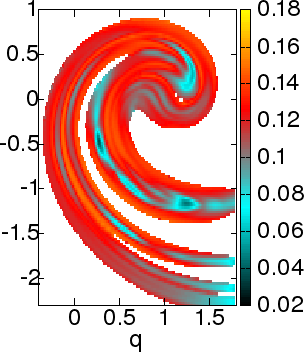}}
\hskip 0.3cm
(c)\scalebox{.65}{\includegraphics{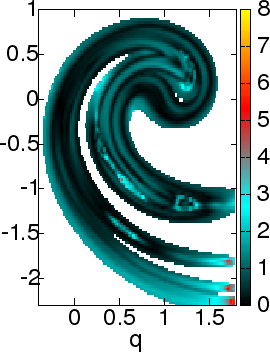}}}
\vskip 0.2cm
\centerline{
(d)\scalebox{.65}{\includegraphics{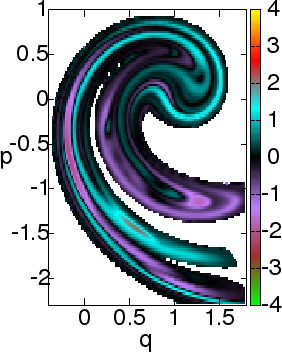}}
\hskip 0.5 cm
(e)\scalebox{.65}{\includegraphics{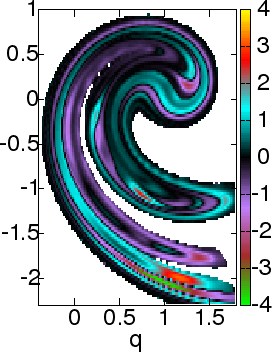}}} 
\caption{(a): Density plot ($2^{14}$ points, each averaged over $10^4$ trajectories)  of the diffusivity for the Ikeda map with noise of amplitude $2\sigma^2=2.5\cdot10^{-3}$, pinned by the final points, after $t=8$ iterations of the map. (b): Distribution of the finite-time Lyapunov exponents, $t=8$. (c): Ratio of  (b) to (a), as defined in~\refeq{ratio}. (d)-(e): Real- and imaginary parts respectively of the second eigenfunction of the Perron-Frobenius operator, divided by the first eigenfunction.}
\label{ikfin}
\end{figure*}

\begin{figure*}[tbh!]
\centerline{
(a)\scalebox{.65}{\includegraphics{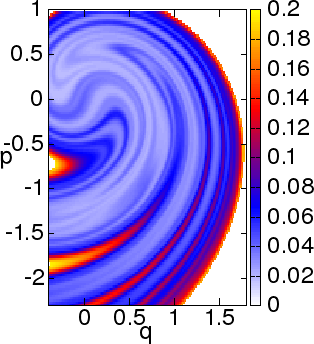}}
\hskip 0.3cm
(b)\scalebox{.65}{\includegraphics{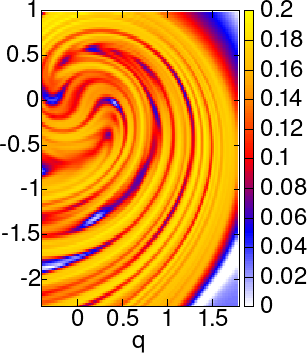}}
\hskip 0.3cm
(d)\scalebox{.65}{\includegraphics{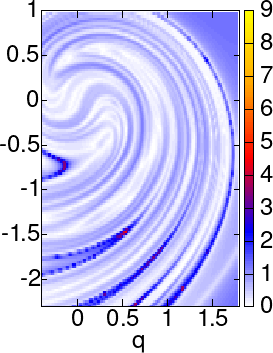}}}
\vskip 0.2cm
\centerline{
(d)\scalebox{.65}{\includegraphics{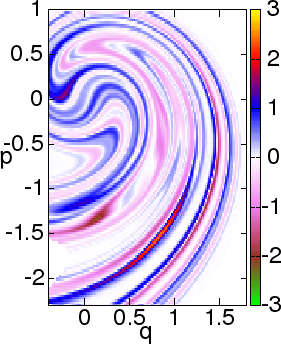}}
\hskip 0.5 cm
(e)\scalebox{.65}{\includegraphics{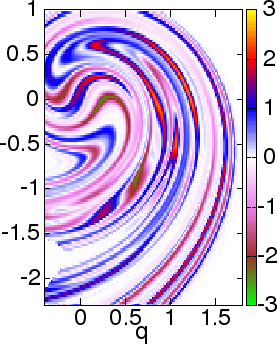}}} 
\caption{(a): Density plot  ($2^{14}$ points, each averaged over $10^4$ trajectories)  of the diffusivity  for the Ikeda map with noise, pinned by the initial points, after $t=8$ iterations of the map. (b): Distribution of the finite-time Lyapunov exponents, $t=8$. (c): Ratio of  (b) to (a) , as defined in~\refeq{ratio}.  (d)-(e): Real- and imaginary parts respectively of the second eigenfunction of the Koopman operator, divided by the first eigenfunction.}
\label{ikin}
\end{figure*}

In order to verify that, the density plots in Figs.~\ref{ikfin}(a)-(b) and~\ref{ikin}(a)-(b) compare two distinct integrated observables, that is the diffusivity $\hat{D}^t(x)$ defined as in Eq.~\refeq{xyDiff}, 
and the finite time Lyapunov exponent, defined in Eq.~\refeq{FTLE}. This time the directly computed
observables are both supported on and striated along  the unstable (Fig.~\ref{ikfin}) and 
stable (Fig.~\ref{ikin}) manifolds respectively, but they share limited similarities.

 In particular, certain features that belong now to the real- now to the imaginary part of the second eigenfunction (divided by the first, that is real valued, Figs.~\ref{ikfin}(d)-(e) and~\ref{ikin}(d)-(e)) may be visible in the profiles of either observable (Figs.~\ref{ikfin}(a)-(b) and~\ref{ikin}(a)-(b)), but not in a consistent manner, as seen for the non-dissipative models of the previous sections.
 
 The ratios $r(x,t)$, as defined in~\refeq{ratio}, between the two integrated observables (Figs.~\ref{ikfin}(c) and~\ref{ikin}(c))
 are also striated along the manifolds, meaning that they are no longer uniform distributions with 
 fluctuations. That indeed gives an additional indication that distinct integrated observables produce 
 different phase-space profiles.    
 
 \section{Summary}
 I have carried out an attempt to take the thermodynamic formalism of chaotic dynamics out of 
 statistical equilibrium. The time evolution of the phase space is treated as 
 a thermodynamic ensemble, and the single chaotic trajectories as its susbsystems.
 The probability for the value of a given observable on a trajectory follows the usual Gibbs expression,
 which however now includes a time-dependent statistical weight for each orbit.
 
 With those premises, the familiar expressions relating R\'enyi (e. g. information) entropy, a free energy (`topological pressure'), and ensemble averages, are extended to chaotic processes that have yet to relax to statistical equilibrium.     
 
 The most notable byproduct of the present construction emerges from the evaluation of the 
 ensemble average of an integrated observable.  According  to the theory, every such expectation 
 value does depend explicitly on a phase-space weight distribution, which in turn depends both on time and,
 crucially, on the initial conditions. That makes average and higher moments not so meaningful by themselves, and so one rather studies the behavior of the full phase-space profiles of an integrated 
 observable, which are found to be determined by the first two eigenfunctions of the
 transport (Perron-Frobenius or Koopman) operator, at an intermediate timescale during relaxation.
 This outcome, and the prediction of a universal behavior for the profile, 
 is obtained with an alternative approach to that of dynamical averages, used for both pointwise-
 and integrated observables in a recent report~\cite{LipShort24}, and thus independently confirms 
 the conclusions of the present work. 
 
 However, the present results  
 are subject to a number of assumptions, and thus limitations. First, the theory
is here formulated in discrete time, and thus an extension to continuous-time flows is in order. Secondly, in order for the observations on the eigenfunctions to apply,
 the dynamics must allow for a
 transport operator with a discrete spectrum, and a spectral gap. Typically, that occurs with a 
 strongly chaotic (`hyperbolic') phase space, or, as seen for the Hamiltonian H\'enon map, a chaotic repeller bearing a small stability island, and an appropriate choice of the functional space. 
 Finally, the universality of the observable phase-space profiles  breaks down
 when the second eigenvalue of the transport operator is a complex conjugate pair instead of a real-valued 
 singlet. This is ascribed to the phase-space squeezing caused by dissipation, manifest for example in strange
 attractors.
 
 The present approach to finite-time thermodynamic formalism proves self-consistent when I formally
 identify the weights of the out-of-equilibrium Gibbs probabilities with phase-space densities, which
 are pushed forward or pulled back by well-known transport operators. Moreover, physically meaningful expressions for the ensemble averages are recovered, and the predictions for the observables in the phase space are corroborated by numerics. Yet, I did not provide a mathematically rigorous theory akin to that already existing for chaotic systems at statistical equilibrium, where, for instance, it is proved that the conventional Gibbs probabilities extremize the topological entropy. 
Future developments of the present theory should then move in that direction, possibly leveraging variational principles for systems out of equilibrium, such as Maximum Caliber~\cite{MaxCal},  besides the mentioned need for a continuous-time formulation.

 \begin{appendix}
 
 \section{Asymptotic limit for the $I_t,{\cal{P}}_t,\langle A^t\rangle$ thermodynamic relation}
 \label{sec_asymp}
  Let us retrieve
 Eq.~\refeq{PAhEq},  relating information entropy, topological pressure, and expectation value of an integrated observable at equilibrium,
 from the limit $t\rightarrow\infty$ of its finite-time, non-equilibrium counterpart~\refeq{I1outeqS} derived in Sec.~\ref{FiniteTime}.
 
An argument may be used from Reference~\cite{BeckSchl} to help determine the asymptotic behavior of our
finite-time thermodynamic relation connecting Shannon information, Gibbs average, and topological pressure.
Take as integrated observable $A^t = \ln |\Lambda(t)|$, the finite-time Lyapunov exponent, as defined in Sec.~\ref{FiniteTime}, Eq.~\refeq{FTLE}.
Consider the partition function
\beq
Z_t(\beta) = -\sum_j \frac{w_j(t)}{|\Lambda_j(t)|^\beta}
\,. 
\label{TopZ}
\eeq
Assume that all the weights $w_j$ are bounded by positive constants $c_1$, and $c_2$ as
\beq
c_1 \leq w_j \leq c_2
\,.
\eeq
That is reasonable, if the map in question is expanding or hyperbolic, since in that case the dynamics is
everywhere unstable and no trajectory/sequence may ever carry infinite statistical weight.
We may then sandwich the partition function~\refeq{TopZ} as
\beq
c_1\sum_j \frac{1}{|\Lambda_j(t)|^\beta} \leq \sum_j \frac{w_j(t)}{|\Lambda_j(t)|^\beta}
\leq c_2\sum_j \frac{1}{|\Lambda_j(t)|^\beta}
\,,
\ee{Zsandw}
and hence
\begin{widetext}
\bea
\nonumber
\frac{1}{t}  \ln c_1\sum_j \frac{1}{|\Lambda_j(t)|^\beta} \leq \frac{1}{t} \ln \sum_j \frac{w_j(t)}{|\Lambda_j(t)|^\beta} 
\leq \frac{1}{t}  \ln c_2\sum_j \frac{1}{|\Lambda_j(t)|^\beta}  \\
\frac{1}{t}  \ln c_1 + \frac{1}{t} \ln \sum_j \frac{1}{|\Lambda_j(t)|^\beta} \leq \frac{1}{t} \ln \sum_j \frac{w_j(t)}{|\Lambda_j(t)|^\beta} 
\leq \frac{1}{t}  \ln c_2 + \frac{1}{t}\ln\sum_j \frac{1}{|\Lambda_j(t)|^\beta}
\,,
\eea
so that, when $t\rightarrow\infty$, both inequalities saturate to yield the same expression 
\beq
\lim_{t\rightarrow\infty}\frac{1}{t}\ln\sum_j \frac{1}{|\Lambda_j(t)|^\beta} = {\cal{P}}(\beta)
\,.
\eeq
This is the topological pressure defined at equilibrium with the unweighed Gibbs probabilities,
whose ensemble constitutes the pointwise natural measure of the dynamical system at hand~\cite{Gaspbook}.
Besides retrieving ${\cal{P}}(\beta)$ from the $t\rightarrow\infty$ limit of  ${\cal{P}}_t(1,\beta)$, the previous analysis tells us that the quantity $Z_t(\beta)$ given by Eq.~(\ref{TopZ}) is 
also bounded as, say, $c'_1\leq Z_t(\beta)\leq c'_2$.
\end{widetext}
Let us now apply the same idea as in~\refeq{Zsandw} to the tilted information $S_t(\beta)$: let
\beq
\chi_1 \leq w_j\ln w_j \leq \chi_2
\,,
\eeq
for some $\chi_1, \chi_2$, and so
\beq
\chi_1\sum_j \frac{1}{|\Lambda_j(t)|^\beta} \leq \sum_j \frac{w_j(t)\ln w_j(t)}{|\Lambda_j(t)|^\beta}
\leq \chi_2\sum_j \frac{1}{|\Lambda_j(t)|^\beta}
\,,
\ee{Ssandw}
that also results in upper and lower bounds for $t\,S_t(\beta)$. 
Then we may also sandwich the ratio $t\,S_t(\beta)/Z_t(\beta)$ we have encountered in Eq.~\refeq{I1outeqS}:
\beq 
 \frac{\chi'_1}{c'_2}  \leq \frac{\sum_j \frac{w_j(t)\ln w_j(t)}{|\Lambda_j(t)|^\beta}}{\sum_j \frac{w_j(t)}{|\Lambda_j(t)|^\beta}}
\leq \frac{\chi'_2}{c'_1}
\,,
\eeq    
with the assumption that $w_j(t)>0$. The previous bounds are constant in time, and thus, including the fractor of $1/t$ originally in Eq.~\refeq{I1outeqS}, we have that
\beq
\lim_{t\rightarrow\infty} \frac{S_t(\beta)}{Z_t(\beta)} =  \lim_{t\rightarrow\infty}  \frac{1}{t} 
 \frac{\sum_j \frac{w_j(t)\ln w_j(t)}{|\Lambda_j(t)|^\beta}}{\sum_j \frac{w_j(t)}{|\Lambda_j(t)|^\beta}} 
 = 0
\,,
\eeq
and Eq.~\refeq{I1outeqS} does reduce to the steady-state thermodynamic relation~\refeq{LyapTherm} linking information entropy, Lyapunov exponent, 
and escape rate (for $\beta=1$). 

\section{Eigenfunction expansion for backward evolution}
\label{BackEvol}
Let me here provide the intermediate steps leading from Eq.~\refeq{ALop} to Eq.~\refeq{Aadjdistrunc}. First, in the left-hand side of Eq.~\refeq{ALop}, 
I expand $[\Lop^\dagger_t w](x)$ in terms of the eigenspectrum of the Koopman operator, to obtain for the expectation $\langle A^t\rangle$:
 \beq
   \langle A^t \rangle_\beta = \frac{\sum_n^\infty \tilde{b}_n e^{-\gamma_n t} \int dx\, \varphi_n(x)  A^t(x) e^{-\beta A^t(x)}}{\sum_n^\infty \tilde{b}_n e^{-\gamma_n t} \int dx\, \varphi_n(x) e^{-\beta A^t(x)}}
 \,,
  \ee{Adualexp} 
  with 
  \beq
  \tilde{b}_n = \int dy\, w(y)\phi_n(y)
  \,.
  \eeq  
As before with the $\varphi_n(x)$, this time the eigenfunctions $\phi_n(x)$ of the Perron-Frobenius operator are hidden in the coefficients $\tilde{b}_n$ of the expansion, and yet they come out when we take a density pinned at a definite point, $w(x)~=~\delta~(x~-~x_t~)$. Then the average of the integrated observable becomes 
the phase-space function 
\begin{widetext}
  \beq
  \langle A^t \rangle_{\beta=0} (x_t) =
    \frac{ \int dx \, A^t(x) \int dy \, \delta(y-f^t(x)) \delta(y-x_t)}{ \int dx\, \int dy\, \delta(y-f^t(x)) \delta(y-x_t)} = \frac{A^t(f^{-t}(x_t))}{|\mathrm{det}\, J^t(f^{-t}(x_t))|\,\mu(\pS(t))}  
   \,.
   \eeq
Specularly to Eq.~\refeq{Atexp}, I obtain for $\hat{A}^{t}(f^{-t}(x_t))$ in terms of the spectral expansion,
\beq
    \hat{A}^{-t}(x_t) :=  \langle A^t\rangle_{\beta=0}(x_t)  =   \frac{\sum_n^\infty \phi_n(x_t) e^{-\gamma_n t} \int dx\, \varphi_n(x)  A^t(x)} {\sum_n^\infty \phi_n(x_t) e^{-\gamma_n t} \int dx\, \varphi_n(x)} 
    \,.
   \ee{Atdualexpx0}
\end{widetext}
The meaning of the previous expression is that the pointwise expectation of any integrated observable pinned by the \textit{arrival} point $x_t$ in the phase space is a superposition of eigenfunctions of the Perron-Frobenius operator. 
The truncation~\refeq{Atgamma1} to the second eigenfunction may also be applied here to Eq.~\refeq{Atdualexpx0}, for timescales of the order $(\gamma_1~-~\gamma_0)^{-1}$, and it results in Eq.~\refeq{Aadjdistrunc}.

 \end{appendix}


\begin{thebibliography}{}
\bibitem{footnote1} According to reference~\cite{BeckSchl}, the quantity to be minimized is the function
$\Psi_t(a,p) = \sum_j p_j\beta E_j + p_j\ln p_j\,$.
\bibitem{footnote2} I assume the flow $f^t$ to be invertible, so there is a $1-1$ correspondence between every $x$ and $f^{-t}(x)$.
\bibitem{footnote3} Here the evolution operators are approximated with matrices, meaning that the numerically computed normalized eigenvectors are determined up to a sign. In any case the $\varphi_n$ must be bi-orthogonal to the $\phi_n$, while the $A^t(f^{-t}(x))$ considered here are positive definite, hence the sign of $\varphi_n$ determines the sign of the integral $ \int dx \, \phi_n(x)  A^t(f^{-t}(x))$ as well.
\end{thebibliography}
\end{document}